\documentclass[12pt]{article}
\usepackage{epsfig}
\usepackage{amsfonts}
\usepackage{amsmath}
\usepackage{amssymb}
\usepackage{graphicx}

\usepackage{color} 
  \newcommand{\beq}{\begin{equation}}
  \newcommand{\eeq}{\end{equation}} 
  \def\nuc#1#2{\relax\ifmmode{}^{#1}{\protect\text{#2}}\else${}^{#1}$#2\fi}
  \def\itnuc#1#2{\setbox\@tempboxa=\hbox{\scriptsize\it #1}
    \def\@tempa{{}^{\box\@tempboxa}\!\protect\text{\it #2}}\relax
    \ifmmode \@tempa \else $\@tempa$\fi}
  
\newcommand{\simge}{\hspace*{0.2em}\raisebox{0.5ex}{$>$}
     \hspace{-0.8em}\raisebox{-0.3em}{$\sim$}\hspace*{0.2em}}
\newcommand{\simle}{\hspace*{0.2em}\raisebox{0.5ex}{$<$}
     \hspace{-0.8em}\raisebox{-0.3em}{$\sim$}\hspace*{0.2em}}

\newcommand{\dslash}[1]{#1 \llap{/\kern-0.5pt}}
\newcommand{\Dslash}[1]{#1 \llap{/\kern+1.2pt}}
\newcommand{\DDslash}[1]{#1 \llap{/\kern+2.3pt}}
\newcommand{\dslashh}[1]{#1 \llap{/\kern+1pt}}

\def\bdm{\begin{displaymath}}
\def\edm{\end{displaymath}}

\begin{document}

\begin{titlepage}

\vspace*{1.5cm}

\begin{center}
{\Large\bf Improved Actions for
\\
\vspace{0.3cm}
Nuclear Effective Field Theories}
\\

\vspace{2.0cm}

{\large \bf U. van Kolck}

\vspace{0.5cm}
{
{\it European Centre for Theoretical Studies in Nuclear Physics and Related Areas (ECT*)
\\
Fondazione Bruno Kessler, 38123 Villazzano (TN), Italy}
\\
\vspace{0.2cm}
{\it Universit\'e Paris-Saclay, CNRS/IN2P3, IJCLab
\\
91405 Orsay, France}
\\
\vspace{0.2cm}
{\it Department of Physics, University of Arizona
\\
Tucson, AZ 85721, USA}
}

\end{center}

\vspace{1.5cm}

\begin{abstract}
Effective field theories have been successful in describing nuclei up to the alpha particle but face significant challenges for larger nuclei due to leading-order instabilities. These issues can be addressed with the introduction of a fake interaction range at leading order, whose effects are compensated for in perturbation theory at higher orders. The calculation of two-body phase shifts and ground-state energies for up to five $^4$He atoms in a theory with only contact interactions shows that, as long as it remains smaller or comparable to the experimental effective range, the fake range does not alter the convergence of the EFT expansion but is often beneficial at the lowest orders. I discuss the implications of this improved-action approach to the ground-state energies of nuclei such as $^6$Li, $^{12}$C, and $^{16}$O.
\end{abstract}

\end{titlepage}

\section{Introduction}
\label{intro}

Long before my first Spring Symposium in Ischia, my colleague Bruce Barrett told me of this wonderful series of meetings organized by the Napoli and Caserta groups. He waxed lyrical about the unique combination of interesting researchers, beautiful locations, and outstanding hospitality. The tradition continues strong, but this time we mourn Bruce. 

Bruce was a gentle ground-breaker, a warm mentor, and a stimulating friend. His work underlies much of current theoretical nuclear physics and thus many of the contributions to the meeting --- including mine. Bruce had an impactful career, with his most influential work being perhaps the No-Core Shell Model (NCSM) \cite{Barrett:2013nh}. This was the first ``{\it ab initio}'' method in nuclear physics that could accommodate interactions derived from effective field theories (EFTs) \cite{Hammer:2019poc}. In the late 2000s, together we formulated the NCSM as a particular regularization of a renormalizable EFT \cite{Stetcu:2006ey} and observed a tendency for clusterization at leading order (LO). It has been shown since then that EFTs describe systems up to the alpha particle well but, at LO in the large cutoff limit, do not produce states of heavier nuclei that are stable against break up into lighter clusters \cite{Contessi:2017rww, Bansal:2017pwn, Yang:2020pgi, Contessi:2025xue}.

Nowadays much of nuclear theory consists of {\it ab initio} calculations with potentials inspired by Chiral EFT. Although phenomenologically successful at high orders, existing chiral potentials do not produce renormalizable amplitudes and are, therefore, not necessarily a model-independent representation of QCD. As any other EFT, Chiral EFT itself is renormalizable when properly formulated \cite{vanKolck:2020llt}, but the importance of three- and more-body forces remains unclear \cite{Yang:2021vxa}. Another nuclear EFT, Pionless EFT, is much simpler and better understood. I use it here to illustrate how the break-up issue can be tackled. The idea \cite{Contessi:2023yoz,Contessi:2024vae,Contessi:2025xue}, which could also be applied to other EFTs, is to use the intrinsic uncertainty of each order to pick an improved LO where nuclei are stable, and then compensate for the small change order by order in the (perturbative) EFT expansion. 

\section{Short-Range EFT}
\label{sec-SREFT}

QCD for light quarks has essentially two scales: $M_{\rm QCD}\sim 1$ GeV, set by its nonperturbative dynamics; and $m_\pi\simeq 140$ MeV, set by the average light-quark mass. One of the most basic observations about nuclei is that they are weakly bound, in the sense that the characteristic momentum of bound nucleons $Q\ll M_{\rm QCD}$ and, at least for light nuclei, even $Q\simle m_\pi$. We can address nuclear dynamics with EFTs built out of color singlets (chiefly nonrelativistic nucleons) with local interactions, constrained only by the symmetries of QCD.

Several nuclear EFTs have been formulated \cite{Hammer:2019poc}, which differ in their breakdown scale $M_{\rm hi}$. They offer a challenge compared to many other EFTs, because LO interactions are singular and must be nonperturbative to generate the bound states and resonances we call nuclei. A model-independent treatment of singular interactions requires regularization with a cutoff $\Lambda$ and subsequent renormalization to render the regularization procedure arbitrary. The $\Lambda$ dependence of interaction strengths (``low-energy constants'', LECs) is adjusted to ensure the order-by-order removal of non-negative powers of $\Lambda$ from observables. Since these powers signal sensitivity to physics at the scale $M_{\rm hi}$, naturalness (for a review, see Ref. \cite{vanKolck:2020plz}) suggests that, barring a fine-tuning, the magnitude of the remaining (``finite'') part of a LEC should be comparable to the cutoff-dependent part when $\Lambda\to M_{\rm hi}$. The exact value is determined from experimental data or matching to an EFT with a higher breakdown scale. The residual dependence of observables on inverse powers of $\Lambda$ is no larger than higher-order contributions as long as $\Lambda\simge M_{\rm hi}$. 

For predictive power subleading orders must be relatively small and amenable to (distorted-wave) perturbation theory. Renormalization in this context yields surprising features, which are most clearly illustrated with Pionless EFT \cite{Hammer:2019poc} (more generally, Short-Range EFT for degrees of freedom other than nucleons), where $M_{\rm hi}\sim R^{-1}$, the inverse range of the potential. In the nuclear case, $R\sim m_{\pi}^{-1}$ and pions can be integrated out along with other mesons and baryons. Analogously to the multipole expansion of classical electromagnetism, the dynamics is described by a series of delta-function interactions with arbitrary number of derivatives and bodies. The universality of this approach means it can be applied to and tested in other large systems, such as $^4$He atomic clusters where $R\sim l_{\rm vdW}$, the van der Waals length. It can also be applied to nuclear systems at the large quark masses more accessible to lattice QCD simulations \cite{Barnea:2013uqa,Bansal:2017pwn}.

The first couple of orders in Short-Range EFT are well understood following the principles of EFT --- renormalization and naturalness --- apart from one fine-tuning \cite{vanKolck:1998bw} needed to achieve $B_2\ll 1/2mR^2$, where $m$ is the particle mass. An expansion in powers of $M_{\rm lo}/M_{\rm hi}$ around the unitarity limit $B_2\to 0$ works well for $Q\sim M_{\rm lo}\equiv \sqrt{2m B_3/3}$ \cite{Konig:2016utl}. At LO there is a parameter-free, nonderivative two-body delta function and a three-body contact interaction, which is needed for renormalization \cite{Bedaque:1998kg} and can be determined from the three-body ground-state energy. LO has a discrete scale invariance that fixes all observables in terms of the three-body LEC (for a review, see Ref. \cite{vanKolck:2017jon}). At relative ${\cal O}(M_{\rm lo}/M_{\rm hi})$, NLO, the two-body scattering amplitude is endowed with the energy dependence associated with the effective range $r_2\sim M_{\rm hi}^{-1}$ in the effective-range expansion (ERE). Renormalization for the four- or more-body problem requires an NLO four-body force \cite{Bazak:2018qnu} and introduces a scale in correlations among four-body observables \cite{Wu:2023mhg}. If we count the two-body scattering length as $|a_2|\sim M_{\rm hi}/M_{\rm lo}^2$, then the LEC of the nonderivative two-body delta function gets an NLO correction $\propto a_2^{-1}$.

Physics at $Q\sim M_{\rm lo}^2/M_{\rm hi}$ can be accounted for by an infrared resummation, when the actual values of the two-body scattering lengths or the deuteron binding energy $B(^2\textrm{H})$ are fitted at LO. The EFT expansion works well in the two-nucleon system up to and including N$^3$LO \cite{Chen:1999tn,Wu:2025XX}, with a breakdown scale of the expected magnitude \cite{Ekstrom:2024dqr,Wu:2025XX}. A similar value for the breakdown scale emerges when the three-nucleon system is considered \cite{Ekstrom:2025ncs}. For $A\ge 3$ ground states, which are sufficiently deep, the Coulomb interaction can be treated perturbatively, appearing at NLO as a first-order perturbation. With the three-body force fixed by the triton binding energy $B(^3\textrm{H})$ \cite{Bedaque:1999ve}, the helion binding energy at LO is\footnote{I denote  by a superscript $^{(n)}$ the sum of contributions up to and including N$^n$LO.} $B^{(0)}(^3\textrm{He})= B(^3\textrm{H})$ and at NLO \cite{Konig:2015aka},
\begin{equation}
B^{(1)}(^3\textrm{He})= (7.62 \pm 0.17) \; {\rm MeV} ,
\end{equation}
$\simle 1.5\%$ away from the observed value. Isospin breaking from the quark mass difference is a higher-order effect \cite{Konig:2015aka}. As $a_2^{-1}$ increases, the first triton excited state becomes \cite{Rupak:2018gnc} the virtual state seen in $n$-$^2$H scattering. At LO, the alpha-particle binding energy $B^{(0)}(^4\textrm{He})/B(^3\textrm{H}) \simeq 4.7$, and scattering-length corrections bring this value within $\simeq$ 5\% of the observed $\simeq 3.3$ ratio \cite{Konig:2016utl}. Resumming these corrections \cite{Contessi:2025xue},
\begin{equation}
B^{(0)}(^4\textrm{He})= (31.8\pm 6.6) \; {\rm MeV}.
\end{equation}
The first alpha-particle excited state is just below the $p$-$^3$H threshold (which at this order is degenerate with $n$-$^3$He), $B^{(0)}(^4\textrm{He}^*)/B(^3\textrm{H}) \simeq 1.002$ \cite{Konig:2016utl}, instead of slightly above as in experiment. At NLO, the experimental value of $B(^4\textrm{He})$ is used to fix the four-body force, and predictions for other quantities need to be computed with Coulomb included. In the simpler case of neutral $^4$He atoms, first NLO results \cite{Lin:2023zqw,Wu:2025XY} indicate a similar level of success as for light nuclei. Low-energy scattering for $A\le 5$ \cite{Schafer:2022hzo,Bagnarol:2023crb} is also well described.

For $A\ge 5$ nucleons, LO tends to have free clusters as ground state \cite{Dawkins:2019vcr}. This feature does not seem to be affected qualitatively \cite{Contessi:2017rww, Bansal:2017pwn, Contessi:2025xue} by a resummation of the relative ${\cal O}(M_{\rm lo}/M_{\rm hi}$) effects associated with the two-body scattering lengths. This is not necessarily a fundamental problem, since break-up energies are usually much smaller than binding energies, which is reflected in cluster substructures. It is in principle possible to start with free clusters at LO and obtain stable states from small changes at NLO. There is, however, the practical issue of how to do distorted-wave perturbation theory starting from free clusters. A way around this problem is described next.

\section{Improved actions}
\label{sec-imp}

It is essential to recall that the values an EFT give for an observable at N$^n$LO come with a relative uncertainty of ${\cal O}(M_{\rm lo}^{n+1}/M_{\rm hi}^{n+1})$. Thus, at a given order one can always include higher-order effects, as long as they do not change the physical content of the order being considered and can therefore be compensated perturbatively at higher orders. In fact, since data used as input to determine the LECs include all orders, it is inevitable that some higher-order effects are included through renormalization conditions. The choice of low-energy observable used to determine a LEC is arbitrary, as the difference will be largely removed at next order, and more subsequently. The simplest example is offered by Short-Range EFT in the two-body sector: the finite part of the LEC for the nonderivative two-body contact interaction can be fixed from either the scattering length or the binding energy, the difference being reduced progressively by the inclusion of interactions accounting for the effective range and ERE shape parameters. The crucial issue is what a ``low-energy observable'' is: a quantity for which the EFT expansion converges. For example, a chiral potential, NNLO$_{\rm sat}$ \cite{Ekstrom:2015rta}, has been fitted to many-body, instead of only few-body, data, which is fine as long as we have, or are gathering, evidence that such dense nuclei are within the realm of the theory. But this might not be so easy to assess, and a more general method is desirable.

We can generalize this method to any small interaction, which we can add to LO. The test of whether the change it brings remains of relative ${\cal O}(M_{\rm lo}/M_{\rm hi})$ is that we must be able to reduce its effects at NLO to a relative ${\cal O}(M_{\rm lo}^2/M_{\rm hi}^2)$, and so on. The interaction strength is limited by this constraint but, because it is superfluous when capturing ${\cal O}(1)$ effects, it does not need to be fitted to data --- it does not bring a new parameter to LO as a change in power counting would. It does neither replace an LO interaction nor add a new one. By capturing some effects within the LO uncertainty band, it can lead to central values for observables that are closer to the prediction from higher orders. However, this method does not affect the breakdown scale $M_{\rm hi}$, which is a physical scale of the underlying theory, and what the EFT expansion converges to does not change. This method is a generalization of the improved actions in lattice quantum field theory \cite{Symanzik:1983dc}, from which we borrow the name. By its nature, the improvement is not unique. (It can also make central values go in the opposite direction within the error band!) If we find that the added interaction cannot be compensated perturbatively at higher orders, it is not acceptable under the existing power counting: either another improvement must be sought or the power counting rethought. Before blaming power counting for the LO instability problem, we try to improve LO so as to produce (shallow) stable states at LO on top of which we can perturb.

A natural improvement is to include in LO some of the energy dependence manifest in the ERE, which is most easily done through the introduction of an auxiliary dimer field \cite{Kaplan:1996nv}. In fact, inclusion of the two-body effective range was considered in Ref. \cite{Beane:2000fi}. However, because $r_2$ is really an NLO effect, one can add to LO just a fraction $x$ of it. A simple two-body toy model \cite{Contessi:2024vae} confirms that as long as $|x|\simle M_{\rm hi}/M_{\rm lo}$, its effects can be compensated perturbatively at higher orders. Central-value errors might decrease ($x>0$) or increase ($x<0$), but $M_{\rm hi}$, estimated via Lepage plots \cite{Lepage:1997cs}, does not change. The limitation on $x$ is related to the intrusion of a second pole at low energies. 

Unfortunately, accounting for energy dependence directly in the interaction poses difficulties in the many-body context. The obvious alternative is to introduce a fake interaction range at LO, since central values for binding energies tend to decrease with $\Lambda$ \cite{Contessi:2017rww, Bansal:2017pwn}, which is the inverse of a distance. For example, we can include two- and/or three-body Gaussians with a range $\tilde{R}$ and adjust the LECs already present at each order to ensure subleading orders remain perturbative. At the two-body level, this induces at improved LO not only an effective range but also a correlated series of shape parameters. Taking $^4$He clusters as an example, we have demonstrated  \cite{Contessi:2023yoz} that fake-range effects can be compensated at NLO for $\tilde{R}\simle \tilde{R}_{\rm c}$, where $\tilde{R}_{\rm c}$ is the value for which the induced effective range equals the physical $r_2$. 

In the atomic $^4$He case, the improvement has few practical advantages, since there is no instability problem at $\tilde{R}=0$. The real challenge is to find an improvement where stability takes place at an $\tilde{R}_{\rm u}<\tilde{R}_{\rm c}$, so there is a range of $\tilde{R}$ values where fake-range effects remain perturbative. We have shown \cite{Contessi:2025xue} that such a range exists for a particular improvement where both two- and three-body forces are improved with three fake ranges tied by a single factor $x$, which measures the departure from fake ranges that produce the correct two-nucleon effective ranges and the value for $B^{(0)}(^4\textrm{He})$ closest to experiment. This particular improvement allows variation in only a narrow window $0.9\simle x\simle 1$. For $x= 1$ we find
\begin{equation}
B^{(1)}(^6\textrm{Li}) = (31.6\pm 0.3) \; {\rm MeV},
\end{equation}
\begin{equation}
B^{(1)}(^{12}\textrm{C}) = (97\pm 5) \; {\rm MeV},
\end{equation}
and
\begin{equation}
B^{(1)}(^{16}\textrm{O}) = (156\pm 20) \; {\rm MeV},
\end{equation}
where errors are estimated from variation in the cutoff. The central values are within $\sim 20\%$ of experiment, consistent with the expected expansion parameter $M_{\rm lo}/M_{\rm hi}\sim 0.5$. 

\section{Conclusion}
\label{sec-conc}

The improvement I described is a proof-of-principle that the LO instability of Pionless EFT can be overcome. However, the relative discrepancy with experiment is increasing with $A$, which could be accidental because of the small number of nuclei investigated so far, or an indication of the breakdown of the theory for sufficiently heavy nuclei. More nuclei and higher orders are needed to increase confidence in the convergence of the expansion. 

Perhaps more effective improvements exist, but the one considered so far brings Short-Range EFT to a form that resembles a potential that has been successfully deployed to the study of near-unitarity systems \cite{Gattobigio:2019omi}. Yet it complies with the EFT tenets of order-by-order renormalizability and power counting, which yield a model-independent, systematic expansion. It suggests Pionless EFT has a chance to provide the basis for an understanding of nuclei based on the discrete scale invariance that governs multi-body dynamics near two-body unitarity.

\section*{Acknowledgments}
I thank the organizers of the 14th Spring Seminar for a most enjoyable meeting and my collaborators, in particular Lorenzo Contessi, for their essential contribution to the work reported here.

\end{document}